\documentclass[pra,aps,eqsecnum,showpacs]{revtex4}

\usepackage{amsmath}
\usepackage{graphicx}
\usepackage{epstopdf}
\usepackage[english]{babel}

\usepackage{color}

\begin{document}
\def\BE{\begin{equation}}
\def\EE{\end{equation}}
\def\BY{\begin{eqnarray}}
\def\BEA{\begin{eqnarray}}
\def\EY{\end{eqnarray}}
\def\EEA{\end{eqnarray}}
\def\L{\label}
\def\nn{\nonumber}
\def\({\left (}
\def\){\right)}
\def\<{\langle}
\def\>{\rangle}
\def\[{\left [}
\def\]{\right]}
\def\o{\overline}
\def\BA{\begin{array}}
\def\EA{\end{array}}
\def\ds{\displaystyle}
\def\c{^\prime}
\def\cc{^{\prime\prime}}

\title{Storage and retrieval of squeezing  in multimode resonant quantum memories}
\author{K.~Tikhonov, K.~Samburskaya, T.~Golubeva, Yu.~Golubev}
\affiliation{St.~Petersburg State University, 198504 St.~Petersburg, Stary Petershof, ul. Ul'yanovskaya, 1, Russia}

\date{\today}
\begin{abstract}

In this article the ability to record, store, and read out the quantum properties of light is studied. The discussion is based on high-speed and
adiabatic models of quantum memory in lambda-configuration and in the limit of strong resonance. We show that in this case the equality between
efficiency and squeezing ratio, predicted by the simple beamsplitter model, is broken. The requirement of the maximum squeezing in the output pulse
should not be accompanied by the requirement of maximum efficiency of memory, as in the beamsplitter model. We have demonstrated a high output pulse
squeezing, when the efficiency reached only about 50\%.

Comprehension of this "paradox"\; is achieved on the basis of mode analysis. The memories eigenmodes, which have an impact on the memory process, are
found numerically.  Also, the spectral analysis of modes was performed to match the spectral width of the input signal to the capacities of the
memories.

\end{abstract}
\maketitle

\section{Introduction}

Quantum memory is considered as one of the major challenge for quantum communication and a key element for quantum computing/simulation schemes based
on linear optics during a past decade \cite{repeaters1,repeaters2,repeaters3,comp1,comp2,appl}. The main objective of quantum memory is to write,
store and read out on demand a light pulse without destroying its quantum state.

To implement the quantum memory, many different protocols were proposed (see reviews \cite{review1,review2}), based on these a number of key
experimental advances have been achieved (see for instance \cite{ex1,ex2,ex3,ex4,ex5,ex6,ex7,ex8}). Recently the interests of community shift to the
protocols ensured not only high efficiency of quantum state storage and retrieval, but also fit for storage of the signals with many degrees of
freedom \cite{Nunn2013,Nunn2008,Simon,high-bandwidth,Sorensen,Tittel,SokolovVasilyev,HSQM1}.

Broadband or spatially multimode schemes provide this multiplexing. Addressing to quantum protocols of long-distant communication, for which memory
cells are the key tool, it is clear that the information capacity of the channel strongly depends on its spectral bandwidth, so that the cell memory
as a spectral filter should satisfy some requirements. On the other hand, the involvement of transverse spatial degrees of freedom of the cell and
"parallelization" of information stream is also an important resource of spatial multiplexing. Assessing the benefits of quantum memories for
production of multiphoton states, authors \cite{Nunn2013} have introduced an parameter $\eta B$, where $\eta$ and $B$ are the efficiency and
time-bandwidth product of the memories, respectively (and $B$ is the product of the acceptance bandwidth of the memories and their coherence
lifetime). Thus, in addition to the high efficiency, we have to be concerned about the spectral properties of the proposed memory protocols. Here we
analyze and compare in details the efficiency along with the spectral bandwidths of two memory protocols.

Well known that if you consider off-resonant interaction between atomic ensemble (prepared in the ground state) and two fields - signal and driving,
than the interaction Hamiltonian for such a system, after elimination of the excited state, can be cast in the form \;\;$ \hat H \sim \hat a \hat
b^{\dag} +h.c.$\;\; In quantum optics this Hamiltonian is well-known as the beamsplitter Hamiltonian. The interface mixes the input atom and light
states as a "beamsplitter" and the "reflection coefficient" of unity corresponds to a perfect state swapping between light and atoms.

However, this ideology not always may be applicable. If for some reason the upper excited state can not be excluded from consideration, such schemes
are not beamsplitter-like, and therefore results proved for beamsplitter-like schemes can not be used to them without further justification.
Moreover, even for the off-resonant models of memory, when one takes into account the spatial aspects of light-matter interaction and considers the
propagation of the fields in thick atomic layer, the direct beamsplitter analogy is also inapplicable.

We consider here two quantum memory protocols based on the resonant interaction between an atomic ensemble and fields - signal and driving. The
schemes differ in duration of interaction that leads to differences in formation of the atomic coherence, on which state of the signal field is
mapped. One of our main issues here is to analyze the quantum efficiency. We want to find out whether this characteristic of memory is as universal
for non beamsplitter-like protocols as for beamsplitter-like ones, i.e. could we predict how well a certain quantum state will be survived in the
memory, if we know only the efficiency of this memory.

To demonstrate our arguments, we consider the storage of squeezing in the schemes of adiabatic and high-speed quantum memory. Thereto we analyze how
the squeezed light from a particular source with the desired properties is mapped on the atomic ensemble and then read out. Squeezing storage was
analyzed in \cite{Dantan}, but as opposed to our consideration the authors there do not touch any spatial aspects of interaction and solved the
problem for the flat input spectrum.

We will also solve the eigenfunction problem for two considered memory protocols. A similar problem was solved for Raman-type protocols
(\cite{Nunn,SokolovVasilyev}), when the transformation is unitary. Based on the analysis of eigenfunctions, we compare the spectral bandwidth of the
memories.

\section{Efficiency and storage of squeezing in beamsplitter like quantum memory}\L{BS}

An ideal quantum memory  (QM) maps the quantum state of a light pulse into a long lived coherence in the atomic
ground state using a 2-photon transition, with the help of a strong classical control field. The signal retrieval,
obtained through the interaction of the atomic coherence with the control field should provide a light pulse with the
same classical and quantum properties as the input pulse. Several criteria have been proposed to measure the quality
of this quantum memory. The first one is the efficiency, which measures the ratio of the input and output field
intensities and uses classical quantities. The second one is the fidelity, $\emph{F}$, with
$\emph{F}=<\psi\mid\rho\mid\psi>$, where $\psi$ is the initial wave vector (pure state) of the light, and $\rho$ is
the density matrix describing the state of light after retrieval. The fidelity is a measure of the conservation of
the quantum features of the signal, and allows characterizing the memory as compared to a classical memory, for which
the fidelity is limited to some maximal value. However, the fidelity depends on the stored state and cannot be
considered as a characterization of the memory device itself. A third benchmark the "TV criterion"\; was introduced
in Ref. \cite{Lam} and involves a classical feature, the transmission (efficiency) but also quantum characterization
with the conditional variance V, and allows to establish quantum bounds, while evaluating the classical
characteristics. In this paper, we will be interested in both the classical and the quantum benchmarks and we will
consider the specific example of squeezed light storage, where several factors can impact the retrieved field.

An intuitive understanding of the connection between the storage of squeezing and the efficiency can be gained from the "beamsplitter" model of a
quantum memory, where the a full writing and reading cycle can be modelled by the transmission through a beamsplitter. Indeed, let us write the
relation between input signal $\hat a_{in} $ and output signal $ \hat a_{out} $ in the form of the beamsplitter transformation:
\BY
&&\hat a_{out }(t)=\sqrt {\cal T} \;\hat a_{in }(t)-\sqrt {1- {\cal T}} \;\hat a_{vac }(t).\L{Eff1}
\EY
Here, the "transmission"\; coefficient $ {\cal T} $ models the efficiency of the quantum memory process. Let us rewrite this in the Fourier domain:
\BY
&&\hat a_{out,\omega}=\sqrt{\cal T} \;\hat a_{in,\omega}-\sqrt {1-{\cal T}} \;\hat a_{vac,\omega}. \L{Eff2}
\EY
Going from the field amplitudes $\hat a=\hat x+i\hat y$ to the fluctuations of the quadrature components $\delta \hat x$ and $\delta\hat y$ we get

\BY
&&\langle\delta\hat x_{out,\omega}\delta\hat x_{out,-\omega}\rangle=  {\cal T} \;\langle\delta\hat x_{in,\omega}\delta\hat x_{in,-\omega}\rangle+
(1-{\cal T}) \;\langle\hat x_{vac,\omega}\hat x_{vac,-\omega}\rangle. \L{Eff3}
\EY
Introducing the squeezing parameters for the input field $ r_{in} (\omega) $ and for the output field $ r_{out} (\omega) $, we obtain
\BY
&&\langle\delta\hat x_{out,\omega}\delta\hat x_{out,-\omega}\rangle=\frac{1}{4}\;e^{\ds -r_{out}(\omega)},\qquad\langle\delta\hat
x_{in,\omega}\delta\hat x_{in,-\omega}\rangle=\frac{1}{4}\;e^{\ds -r_{in}(\omega)}.\L{Eff4}
\EY
For the vacuum field amplitude
\BY
&&\langle\delta\hat x_{vac,\omega}\delta\hat x_{vac,-\omega}\rangle=\frac{1}{4}.\L{Eff5}
\EY
As a result we get the equation
\BY
&&1-e^{\ds -r_{out}(\omega)}={\cal T}\[1-e^{\ds -r_{in}(\omega)}\].\L{Eff6}
\EY
We can conclude that the efficiency allows a full characterization of the quantum memory for squeezing transfer if
the beamsplitter model is applicable. If the efficiency is close to unity, then the light squeezing is well preserved
in the read-out pulse.

In this article, we will consider two cases of resonant quantum memory, namely the adiabatic QM \cite{Adiabatic} and
high-speed QM \cite{HSQM1,HSQM2}. They are based on the ensemble of three-level atoms in $ \Lambda $-configuration
(see Fig. 1).
\begin{figure}
 \centering
 \includegraphics[height=35mm]{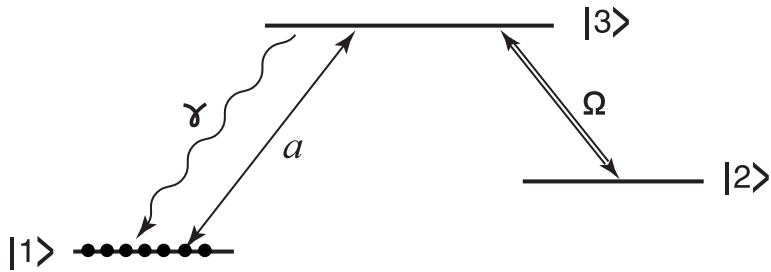}
 \caption{Three level atomic system interacting with driving field $\Omega$ and signal field $a$.}
 \label{fig1}        
\end{figure}
Comparing with the non-resonant schemes, here the population of the upper level is non negligible, which leads to
losses in the signal field during the light-matter interaction as well as during the storage time. As a result, the
analogy between quantum memory and beamsplitter vanishes, and the ratio (\ref{Eff6}) can be not satisfied.

\section{Input-output relation for signal field in three-level resonant quantum memories}\L{NCA}

In this paper, we consider an ensemble of three-level atoms in a $\Lambda$-configuration (Fig. 1) that will be used to store temporal and spatial
multimode quantum fields. The atoms resonantly interact with two electromagnetic fields, a signal field $E_s$ and a driving field $E_d$, that connect
the two atomic ground states to the excited state. The driving field is a strong classical field propagating as a plane wave, while the signal field
is a weak quantum field with a transverse structure. During the writing process both fields interact with the medium simultaneously.

The dynamics of such a system is described by a set of equations for connecting the signal field amplitude $\hat a$ that is going to be recorded in
the atomic medium, operator $\hat b$, associated with the atomic coherence on the transition $|1\rangle - |2\rangle$, which arises in the writing
process, and operator $\hat c$, associated with an atomic coherence on the transition $|1\rangle - |3\rangle$ \cite{Gorshkov_free}:
\BY
&& \frac{\partial}{\partial z} \hat a(z,t;\vec q)=- g\sqrt N \;\hat c(z,t;\vec q),\L{ }\L{1}\\
 &&\frac{\partial}{\partial t}\hat c(z,t;\vec q)=-\gamma\hat c(z,t;\vec q)+   g\sqrt N \;\hat a(z,t;\vec q)
 +\Omega \hat b(z,t;\vec q)+\sqrt{2\gamma}\;\hat\xi(z,t;\vec q),\L{2}\\
 && \frac{\partial}{\partial t}\hat b(z,t;\vec q)= - \Omega^\ast\hat c(z,t;\vec q)\L{3},
\EY
Here $N$ is a mean density of atoms, $g$ is the coupling constant between atoms and field on the $|1\rangle -
|3\rangle$ transition, $\Omega$ is a Rabi frequency of the driving field pulse with a rectangular profile, $\gamma$
is the spontaneous relaxation rate of upper level. These equations are written in the Fourier domain relative to the
transverse co-ordinates ($\vec r=(\vec\rho,z)$ and $\vec\rho\to\vec q$). The Langevin source $\hat\xi(z,t;\vec q )$
is a due to the decay of optical atomic coherence and is assumed to fulfill the following conditions (derived in the
same approximations as the equations (\ref{1})-(\ref{3})):
\BY
&&\langle\hat \xi(z,t;\vec q)\rangle=0,\qquad\langle\hat \xi(z,t;\vec q)\hat \xi^\dag(z^\prime,t^\prime;\vec
q^\prime)\rangle=\delta(t-t^\prime)\delta(z-z^\prime)\delta^2(\vec q-\vec q^\prime),\qquad\langle\hat \xi^\dag(z,t;\vec q)\hat
\xi(z^\prime,t^\prime;\vec q^\prime)\rangle=0.\L{4}
\EY
Amplitudes $\hat a$, $\hat b$, and $\hat c$ have to obey the canonical commutation relations:
\BY
&&\[\hat a(z,t;\vec q ),\hat a^\dag(z,t^\prime;\vec q^\prime )\]=\delta(t-t^\prime)\delta^2(\vec q-\vec q^\prime),\\\
&&
\[\hat b(z,t;\vec q ),\hat b^\dag(z^\prime,t;\vec q^\prime )\]=\[\hat c(z,t;\vec q ),\hat c^\dag(z^\prime,t;\vec
q^\prime )\]=\delta(z-z^\prime)\delta^2(\vec q-\vec q^\prime).
\EY

 We will compare the dynamics of the atomic and field variables in two limits: first, when the signal
and driving field assumed to be very short pulses, much shorter than the excited state lifetime $T\ll \gamma^{-1}$,
so that we can neglect the spontaneous emission during the writing process (high-speed quantum memory), and, vice
versa, when $T\gg \gamma^{-1}$ and we can neglect the time derivative of the optical coherence $\hat c$ (adiabatic
quantum memory).

\subsection{Adiabatic Quantum Memory}\L{NCA_Ad}
The adiabatic approximation can be applied if the evolution time of the system greatly exceeds the life-time of atomic state $|3\rangle$. In this
case  we can set $\partial \hat c(z,t;\vec q)/\partial t = 0$ in (\ref{2}). Then  equations (\ref{1})-(\ref{3}) can be written as
\cite{Adiabatic,Gorshkov_free}:
\BY
&& \frac{\partial}{\partial z} \hat a(z,t;\vec q)=- C_1 \;\hat a(z,t;\vec q)
 -C\; \hat b(z,t;\vec q)-{2g\sqrt{N/\gamma}}\;\hat\xi(z,t;\vec q ),\L{Ad1}\\
 && \frac{\partial}{\partial t}\hat b(z,t;\vec q)=
 -C_2\; \hat b(z,t;\vec q)- C^\ast\;\hat a(z,t;\vec q)-{2\Omega^\ast}/{\sqrt{\gamma}}\;\hat\xi(z,t;\vec q )\L{Ad2},
\EY
In Eqs. (\ref{Ad1})-(\ref{Ad2}) we have introduced the following notation for the constants:
\BY
&& C_1=\frac{2g^2N}{\gamma},\qquad C_2=\frac{2|\Omega|^2}{\gamma},\qquad C=\frac{2g\sqrt
N\;\Omega}{\gamma}\qquad\(C_1C_2=|C|^2\),\qquad\Omega=|\Omega|\;e^{\ds i\varphi}.\L{Ad3}
\EY
For simplicity, in the following we will assume $\varphi=0$.

Let us now introduce dimensionless variables for space and time:
\BY
&&\tilde z=C_1z ,\qquad \tilde t=C_2t.\L{Ad4}
\EY
As shown in \cite{Adiabatic} using Laplace transformation, one can get a general solution of this system in the form
\BY  &&\hat { a}(\tilde z,\tilde t;\vec q)=  \int_0^{\tilde t} d\tilde t^\prime \hat { a}(0,\tilde t^\prime;\vec
q)\;G_{aa}(\tilde z,\tilde t-\tilde t^\prime)-p\;\int_0^{\tilde z} d\tilde z^\prime
\hat { b}(\tilde z^\prime,0;\vec q)\;G_{ba}(\tilde z-\tilde z^\prime,\tilde t)+\hat D_a(\tilde z,\tilde t;\vec q),\L{Ad5}\\
 &&\hat { b}(\tilde z,\tilde t;\vec q)=  \int_0^{\tilde z} d\tilde z^\prime\hat { b}(\tilde z^\prime,0;\vec q)\;
 G_{bb}(\tilde z-\tilde z^\prime,\tilde t)-\frac{1}{p}\;\int_0^{\tilde t} d\tilde t^\prime
 \hat { a}(0, \tilde t^\prime;\vec q)\;G_{ab}(\tilde z,\tilde t-\tilde t^\prime)+\hat D_b(\tilde z,\tilde t;\vec q).\L{Ad6}\\
&&p=\sqrt{\frac{C_2}{C_1}}={\frac{|\Omega|}{g\sqrt N}}.\L{Ad7}
\EY

The kernels are expressed with the modified Bessel functions $I_{0}$ and $I_{1}$ as
\BY  &&G_{aa}(\tilde z,\tilde t)= \[e^{\ds - \tilde z}\delta(\tilde t)+ e^{\ds  - \tilde t-\tilde z}\sqrt{{\tilde
z}/{ \tilde t}}\;I_1\(2\;\sqrt{
\tilde t\tilde z}\)\]\Theta(\tilde t),\L{Ad8}\\
 &&G_{bb}(\tilde z,\tilde t)=  \[e^{\ds -\tilde t}\delta(\tilde z)+e^{\ds -\tilde t-\tilde z}\sqrt{{ \tilde t}/{\tilde z}}
  \;I_1\(2\sqrt{\tilde t \tilde z}\)\]\Theta(\tilde t),\L{Ad9}\\
&&G_{ba}(\tilde z,\tilde t)=G_{ab} (\tilde z,\tilde t)=e^{\ds - \tilde t- \tilde z} I_0\(2\sqrt{ \tilde t \tilde
z}\)\Theta(\tilde t).\L{Ad10}
\EY
Here $\Theta(\tilde t)$ is a window function: $\Theta(\tilde t) = 1$ for $0 < \tilde t < \tilde T$ and equals zero
otherwise, where $\tilde T$ is the interaction time ($\tilde T=\tilde T_W$ for the writing stage and $\tilde T=\tilde
T_R$ for the read out). The operators $\hat D_a(\tilde z,\tilde t;\vec q)$ and $\hat D_b(\tilde z,\tilde t;\vec q)$
are functions of of the Langevine source operator $\hat\xi(\tilde z,\tilde t;\vec q)$:
\BY
&& \hat D_a(\tilde z,\tilde t;\vec q)=\int_0^{\tilde t}\int_0^{\tilde z} d\tilde t^\prime d\tilde z^\prime \;\hat
\xi(\tilde z^\prime,\tilde t^\prime;\vec q)\( -\sqrt{2 C_1}\; G_{aa}(\tilde z-\tilde z^\prime,\tilde t-\tilde
t^\prime)+
p\sqrt{2 C_2}\; G_{ba}(\tilde z-\tilde z^\prime,\tilde t-\tilde t^\prime) \),\\
&& \hat D_b(\tilde z,\tilde t;\vec q)=\int_0^{\tilde t}\int_0^{\tilde z} d\tilde t^\prime d\tilde z^\prime \;\hat
\xi(\tilde z^\prime,\tilde t^\prime;\vec q)\( -\sqrt{2 C_2}\; G_{bb}(\tilde z-\tilde z^\prime,\tilde t-\tilde
t^\prime)+\frac{\sqrt{2 C_1}}{p}\; G_{ab}(\tilde z-\tilde z^\prime,\tilde t-\tilde t^\prime) \).
\EY
It is important to note that the noise operators $\hat D_a$ and $\hat D_b$ are independent of the initial conditions.

The solutions given by Eqs. (\ref{Ad5}) and (\ref{Ad6}) can be first used to model the writing process. For this we
should take into account the initial conditions for operator $\hat a$: $\hat a^W(0,\tilde t;\vec q)=\hat
a_{in}(\tilde t;\vec q)$, which means that a signal field in some specific quantum state enters the cell at $\tilde
z=0$. Before the writing process, all the atoms are in the lower level $|1\rangle$, and the initial state of the
coherence operator $\hat b^W(\tilde z,0;\vec q)$ is the vacuum state. Equations (\ref{Ad5}) and (\ref{Ad6}) can then
be written as
\BY  &&\hat { a}^W(\tilde z,\tilde t;\vec q)=  \int_0^{\tilde t} d\tilde t^\prime \hat { a}_{in}(\tilde t^\prime;\vec
q)\;G_{aa}(\tilde z,\tilde t-\tilde t^\prime)-p\int_0^{\tilde z} d\tilde z^\prime
\hat { b}^W(\tilde z^\prime,0;\vec q)\;G_{ba}(\tilde z-\tilde z^\prime,\tilde t)+\hat D_a(\tilde z,\tilde t;\vec q),\L{Ad11}\\
 &&\hat { b}^W(\tilde z,\tilde t;\vec q)=  \int_0^{\tilde z} d\tilde z^\prime\hat { b}^W(\tilde z^\prime,0;\vec q)\;
 G_{bb}(\tilde z-\tilde z^\prime,\tilde t)-\frac{1}{p}\;
 \int_0^{\tilde t} d\tilde t^\prime  \hat { a}_{in}(\tilde t^\prime;\vec q)\;G_{ab}(\tilde z,\tilde t-\tilde t^\prime)+
 \hat D_b(\tilde z,\tilde t;\vec q).\L{Ad12}
\EY
In Eq. (\ref{Ad12}), the second term is the memory term; it shows the creation of the atomic coherence $\hat {b}^W$ due to the interaction with the
input signal $\hat a_{in}$ and driving field $\Omega$, and corresponds to the storage of the input field in the atomic coherence, while the first
term corresponds to the evolution of the initial atomic coherence $\hat {b}^W (z,t=0;\vec q)$ interacting with the driving field and the third term
to added noise. For $z=L$, equation (\ref{Ad11}) gives the output signal field, the first term corresponds to the leakage of the signal field, the
second one to the generation of a field in the mode of the signal field from the interaction of the initial atomic coherence $\hat { b}^W (z,t=0;\vec
q)$ with the driving field and the third one to added noise.

Since we use the Heisenberg formalism, we deal with operators evolving in time and space and acting on fixed state vector. The state vector of the
full system is determined by the states of subsystems - the signal field, the atomic coherence, and Langevin noise source $ \hat \xi $. At the
initial time, i.e. before the beginning of the writing procedure, all of these subsystems are independent, and the state vector $ | \psi \rangle $ is
factorized:
\BY
&&|\psi^W\rangle={|\psi^W\rangle}_a{|\psi^W\rangle}_b{|\psi^W\rangle}_{\xi}= {|\psi^W\rangle}_a{|vac\rangle}_b {|vac\rangle}_{\xi},\L{Ad13}
\EY
where ${|\psi^W\rangle}_a$ is the initially prepared signal field state, ${|\psi^W\rangle}_b$ is the initial state of the atomic coherence (vacuum
state), and ${|\psi^W\rangle}_{\xi}$ is the state of the Langevin noise source $\hat\xi$ (vacuum state).

At the end of the writing stage ($\tilde t=\tilde T_{W}$), the signal field is vacuum, the driving field is zero, and
we assume that the coherence $\hat { b}^W (\tilde z,\tilde t=\tilde T_{W};\vec q)$ does not decay with time.

Turning to the analysis of the reading stage, we use again the Heisenberg-Langevin equations (\ref{Ad5}) and
(\ref{Ad6}), with different initial conditions and different initial state vector. We assume that there is no input
field at the entrance of the cell and the coherence is $\hat b^R(\tilde z,0;\vec q)=\hat b^W(\tilde z,\tilde T_W;\vec
q)$. The state vector of the full system before reading is also factorized:
\BY
&&|\psi^R\rangle={|\psi^R\rangle}_a{|\psi^R\rangle}_b{|\psi^R\rangle}_{\xi}= {|vac\rangle}_a{|\psi^R\rangle}_b{|vac\rangle}_{\xi},\L{Ad14}
\EY
Let us derive the signal field operator during the reading process:
\BY  &&\hat { a}^R(\tilde z,\tilde t;\vec q)=  \int_0^{\tilde t} d\tilde t^\prime \hat { a}^R(0,\tilde t^\prime;\vec
q)\;G_{aa}(\tilde z,\tilde t-\tilde t^\prime)-p\;\int_0^{\tilde z} d\tilde z^\prime \hat { b}^W(\tilde
z^\prime,\tilde T_W;\vec q)\;G_{ba}(\tilde z-\tilde z^\prime,\tilde t)+\hat D_a(\tilde z,\tilde t;\vec q).\L{Ad15}
\EY
In this equation, the second term shows to the generation of a signal field through the interaction of the atomic coherence $\hat { b}^W(z,T_W;\vec
q)$ with the control field and corresponds to the memory read-out. The first term gives the interaction of the input (vacuum) field with the atoms
and the control field. The last term corresponds to added noise from the Langevine reservoir.

Substitute (\ref{Ad12}) into (\ref{Ad15}) we obtain:
\BY  &&\hat { a}^R(\tilde z,\tilde t;\vec q)=  \int_0^{\tilde t} d\tilde t^\prime \hat { a}^R(0,\tilde t^\prime;\vec
q)\;G_{aa}(\tilde z,\tilde t-\tilde t^\prime)+\int_0^{\tilde T_W}\int_0^{\tilde z}  d\tilde t^\prime d\tilde z^\prime
\hat
{ a}_{in}(\tilde t^\prime;\vec q)\;G_{ab}(\tilde z^\prime,\tilde T_W-\tilde t^\prime)\;G_{ba}(\tilde z-\tilde z^\prime,\tilde t) \nn \\
&&-p\;\int_0^{\tilde z} \int_0^{\tilde z^\prime} d\tilde z^\prime d\tilde z^{\prime\prime} \hat { b}^W(\tilde
z^{\prime\prime},0; \vec
q)\;G_{bb}(\tilde z^\prime-\tilde z^{\prime\prime},\tilde T_W)\;G_{ba}(\tilde z-\tilde z^\prime,\tilde t)  \nn\\
&&  -p\;\int_0^{\tilde z} d\tilde z^\prime \hat D_b(\tilde z^\prime,\tilde T_W;\vec q)\;G_{ba}(\tilde z-\tilde
z^\prime,\tilde t) +\hat D_a(\tilde z,\tilde t;\vec q),\L{Ad16}
\EY
Now let us make averaging of this expression over the vacuum subsystems. Then instead of the canonical amplitude
$\hat { a}^R(\tilde z,\tilde t;\vec q)$  the new one arises:
\BY
&&\hat { A}(\tilde z,\tilde t;\vec q)\equiv\langle {vac}^R|\langle{vac}^W| \hat { a}^R(\tilde z,\tilde t;\vec
q)|{vac}^W\rangle|{vac}^R\rangle ,\L{Ad17}
\EY
where $ |{vac}^W\rangle={|vac\rangle}_b {|vac\rangle}_{\xi}$ are  vacuum state vectors before writing and $
|{vac}^R\rangle={|vac\rangle}_a {|vac\rangle}_{\xi} $ are vacuum state vectors before reading. Further we shall call
operator $\hat { A}(\tilde z,\tilde t;\vec q)$ the \emph{non-canonical amplitude} (NCA), because the corresponding
commutation relations have acquired the non-canonical form.

Let us emphasize that we do not average over the complete wave function, but only over those subsystems that are in
the vacuum state. In this case, the quantum nature of the input signal is taken into account, and we can keep some of
the quantum properties of the output signal field. As a result, all the terms except the second one in (\ref{Ad16})
turn out equal to 0:
\BY
&&\hat { A}^R(\tilde L,\tilde t;\vec q)=   \int_0^{\tilde T_W} d\tilde t^\prime  \hat { a}_{in}(\tilde T_W-\tilde
t^\prime;\vec q)\;G(\tilde t,\tilde t^\prime).\L{Ad20}
\EY
We thus cast an integral transformation of the input field operator into the operator of the output field retrieved in the reading process.

For the forward read-out (when the reading driving field propagates in the same direction as the writing driving
field), the kernel $G(\tilde t,\tilde t^\prime)$ reads:
\BY
&&G(\tilde t,\tilde t^\prime)=\int_0^{\tilde L} d\tilde z^\prime G_{ab}(\tilde z^\prime,\tilde
t^\prime)\;G_{ba}(\tilde L-\tilde z^\prime,\tilde t).\L{Ad21}
\EY
For the backward read-out (when these two driving fields propagate in opposite directions) the transformation keeps
the same form (\ref{Ad20}) with the kernel
\BY
&&G(\tilde t,\tilde t^\prime)=\int_0^{\tilde L} d\tilde z^\prime G_{ab}(\tilde z^\prime,\tilde
t^\prime)\;G_{ba}(\tilde z^\prime,\tilde t) ,\L{Ad22}
\EY
where $G_{ab}$ and $G_{ba}$ are given by Eq. (\ref{Ad10})
Although the introduced partially averaged operators do not convey all the quantum properties of the field (because, for example, they do not obey
canonical commutation relations that can be easily seen from comparison of the canonical amplitude (\ref{Ad16}) with non-canonical one (\ref{Ad20})),
they will be very useful in the following. For example, for the $x$-quadrature of the retrieved field  $\hat a^R$,
\BY
&& \langle :\hat x_a(\tilde z, \tilde t;\vec q)\hat x_a(\tilde z^\prime, \tilde t^\prime;\vec q):\rangle= \langle :
\hat X_A(\tilde z, \tilde t;\vec q)\hat X_A(\tilde z^\prime, \tilde t^\prime;\vec q) : \rangle, \qquad \hat A^R=\hat
X_A+i \hat Y_A, \qquad \hat a^R =\hat x_a+i \hat y_a,
\EY
where notation $\langle :\ldots:\rangle$ stands for normal ordering of the operators.

\subsection{High-speed quantum memory\L{A}}
In the high-speed model of quantum memory the signal and driving pulses are assumed to be much shorter than the excited state lifetime $\gamma^{-1}$,
so that we can neglect spontaneous emission during the interaction processes, which eliminates a source of fluctuations and dissipation in (\ref{2}).
Then the equations for signal field amplitude $a(z,t;\vec q)$, atomic coherence between low levels $b(z,t;\vec q)$, and optical coherence between
levels $|1\rangle$ and $|3\rangle$ $c(z,t;\vec q)$ read \cite{HSQM1}:
\BY
&& \frac{\partial}{\partial z} \hat a(z,t;\vec q)=- g\sqrt N \;\hat c(z,t;\vec q),\L{HS24}\\
 &&\frac{\partial}{\partial t}\hat c(z,t;\vec q)=  g\sqrt N \;\hat a(z,t;\vec q)
 +\Omega \hat b(z,t;\vec q),\quad\;\;\;\L{HS25}\\
 && \frac{\partial}{\partial t}\hat b(z,t;\vec q)= - \Omega^\ast\hat c(z,t;\vec q)\L{HS26},
\EY
The notation here is the same as in (\ref{1})-(\ref{3}).

Eqs.~(\ref{HS24})-(\ref{HS26}) can be solved in the general form by using the Laplace transform. Details of the procedures are discussed in Appendix
A. As a result, applying the same technique of averaging over the vacuum subsystems, we can derive the solution for retrieval field formally in the
same form as for adiabatic case:
\BY
&&\hat { A}^R(\tilde L,\tilde t;\vec q)=   \int_0^{\tilde T_W} d\tilde t^\prime  \hat { a}_{in}(\tilde T_W-\tilde
t^\prime;\vec q)\;G(\tilde t,\tilde t^\prime),\L{HS20}
\EY
Here the kernel $G(\tilde t,\tilde t^\prime)$ is given by
\BY
&&G(\tilde t,\tilde t^\prime)=\frac{1}{2}\int_0^{\tilde L} d\tilde z^\prime G_{ab}(\tilde z^\prime,\tilde t^\prime)\;
G_{ba}(\tilde L-\tilde z^\prime,\tilde t)\;\;\;\mbox{ - for the forward retrieval}\L{HS21}\\
&&G(\tilde t,\tilde t^\prime)=\frac{1}{2}\int_0^{\tilde L} d\tilde z^\prime G_{ab}(\tilde z^\prime,\tilde
t^\prime)\;G_{ba}(\tilde z^\prime,\tilde t)\;\;\;\mbox{ - for the backward retrieval},\L{HS22}
\EY
and functions $G_{ab}$ and $G_{ba}$ now read
\BY
&& G_{ab}(\tilde z,  \tilde t)=G_{ba}(\tilde z,\tilde t)=\int_0^{\tilde t}d\tilde t^\prime
g_{ab}(\tilde z,\tilde t-\tilde t^\prime,)g_{ab}^\ast(\tilde z,\tilde t^\prime),\L{HS27}\\
&&g_{ab}(\tilde z, \tilde  t)=e^{\ds-i  \tilde t}\; J_0\(\sqrt{  \tilde t  \tilde z}\)\Theta( \tilde  t),\L{HS28}
\EY
 where $J_{0}$ is the Bessel functions of the first kind of the zero order. Here and below when we discuss
high-speed QM we use dimensionless variables $\tilde t$ and $\tilde z$ given by
\BY
&&  \tilde t=\Omega t,\qquad \tilde z=  {2g^2N}z/{\Omega}.\L{HS29}
\EY

\section{Comparison of the quantum memory performance for pulse energy and for squeezing}

In this section we want to compare the results of two mental experiments obtained for both adiabatic and high-speed
memory models: an experiment to study the efficiency of QM and experiment for the storage of squeezed light. As shown
in Section \ref{BS} the results of such experiments are uniquely related for the case of the beamsplitter-type
interaction. We will consider whether they related in some way in our models.

The study of the quantum efficiency of the QM implies the following scheme: signal and control pulses with some
defined profiles both are incident on the input face of the atomic cell. Propagating through the cell, these fields
generate a coherence between the two lower levels with a certain spatial distribution, which depends both on the
nature of light-matter interaction and on the duration of this interaction. After some storage time, the retrieved
control pulse  is sent to the input or output cell face and triggers the emission at the signal frequency, which is
collected by a photodetector. One compares the total photon numbers in reconstructed signal and in the input pulse.

Analytically, the quantum efficiency can be read as the following ratio of the averages:
\BE
{\cal E}_{\vec q}=\int_0^{\tilde T_R      } d\tilde t\langle \hat a^\dag_{out}(\tilde t;\vec q)\hat a_{out}(\tilde
t;\vec q)\rangle\(\int_0^{\tilde T_W} d\tilde t\langle \hat a^\dag_{in}(\tilde t;\vec q)\hat a_{in}(\tilde t; \vec q
)\rangle\)^{-1 }.\L{Ex1}
\EE
Then taking into account (\ref{Ad20}), one can obtain for the arbitrary input pulse
\BY
&&{\cal E}_{\vec q}= \L{4.2} \\
&&=\(\int_0^{\tilde T_R}d\tilde t\int_0^{\tilde T_W} d\tilde t^\prime\int_0^{\tilde T_W} d\tilde t^{\prime\prime}
\langle\hat { a}_{in}^{\dag}(\tilde T_W-\tilde t^\prime;\vec q)\hat { a}_{in}( \tilde T_W-\tilde
t^{\prime\prime};\vec q)\rangle\; G(\tilde t,\tilde t^\prime) G(\tilde t,\tilde t^{\prime\prime})\)\(\int_0^{\tilde
T_W}d\tilde t \langle\hat {a}_{in}^{\dag}(\tilde t;\vec q)\hat {a}_{in}(\tilde t;\vec q)\rangle\)^{-1}.\nn
\EY
Choosing a rectangular signal profile, we have to neglect the time dependencies  of the input correlation functions
\BY
 &&\langle\hat { a}_{in}^{\dag}(\tilde T_W-\tilde t^\prime;\vec q)\hat { a}_{in}( \tilde T_W-\tilde
t^{\prime\prime};\vec q)\rangle=\langle\hat {a}_{in}^{\dag}(\tilde t;\vec q)\hat {a}_{in}(\tilde t;\vec
q)\rangle=\langle\hat {a}_{in}^{\dag}(\vec q)\hat {a}_{in}(\vec q)\rangle.
\EY
Then the formula for the efficiency turns out to be essentially simpler
\BY
&&{\cal E}_{\vec q}= \frac{1}{\tilde T_W} \int_0^{\tilde T_R}d\tilde t\(\int_0^{\tilde T_W} d\tilde t^\prime G(\tilde
t,\tilde t^\prime)\)^2.\L{Ex2}
\EY
One can see, for different components of $\vec q$ the efficiencies are equal.

In the second case, we assume that the input pulse is squeezed. Let us assume that pulsed light with given quantum properties is obtained from a
steady source of squeezed light in which a pulse has been cut. It is this kind of light source that was used in the experiments on quantum memory for
squeezed states \cite{Lvovsky}.

Here, we consider two examples of pulsed quantum light generation when the single pulse is obtained by cutting the
time of radiation 1) of single-mode phase-locked sub-Poissonian laser, which emits a squeezed light in a close to
coherent state \cite{SPL}, and 2) of the degenerate optical parametric oscillator (DOPO), operating in the
subthreshold regime \cite{DOPO1, DOPO2}. To assess the conservation of the quantum light properties, we will compare
the quantum properties of the read-out pulse with the original one.

Let us introduce the \emph{degree of light squeezing}  ${\cal S}_{in,\omega,\vec q}$ and ${\cal S}_{out,\omega,\vec q}$ for the input and output
pulses, respectively:
\BY
&&{\cal S}_{out,\tilde \omega,\vec q}=e^{\ds -r_{out}(\tilde \omega,\vec q)},\qquad{\cal S}_{in,\tilde \omega,\vec
q}= e^{\ds -r_{in}(\tilde \omega,\vec q)},\qquad\tilde \omega=\omega/\Omega.\L{Ex4}
\EY
We assess the ratio of these parameters to compare squeezing in the output pulse with squeezing in the input one.
Taking into account Eq. (\ref{Eff4}) this ratio can be derived in the form
\BY
&&\frac{1-{\cal S}_{out,\tilde \omega,\vec q}}{1-{\cal S}_{in,\tilde \omega,\vec q}}=\frac{\langle:\delta\hat
x_{out,\tilde \omega,\vec q}\;\delta\hat x_{out,-\tilde \omega,-\vec q}:\rangle}{\langle:\delta\hat x_{in,\tilde
\omega,\vec q}\;\delta\hat x_{in,-\tilde \omega,-\vec q}:\rangle}.\L{4.4}
\EY
According to (\ref{HS20})  input and output correlation functions in time domain are connected with the expression:
\BY
&& \langle:\delta\hat x^R(\tilde L,\tilde t,\vec q)\;\delta\hat x^R(\tilde L,\tilde t^\prime,\vec
q^\prime):\rangle=\int_0^{\tilde T_W} d\tilde t_1 \int_0^{\tilde T_W} d\tilde t_2\langle:\delta\hat x_{in}(\tilde
T_W-\tilde t_1,\vec q)\;\delta\hat x_{in}(\tilde T_W-\tilde t_2,\vec q^\prime):\rangle G(\tilde t,\tilde t_1)G(\tilde
t^\prime,\tilde t_2).
\EY
Passing to the Fourier domain one can obtain
\BY
&& \langle:\delta\hat x_{out, \tilde \omega,\vec q}\;\delta\hat x_{out, \tilde \omega^\prime,\vec
q^\prime}:\rangle=\int_0^{\tilde T_W} d\tilde t_1 \int_0^{\tilde T_W} d\tilde t_2\langle:\delta\hat x_{in}(\tilde
T_W-\tilde t_1,\vec q)\;\delta\hat x_{in}(\tilde T_W-\tilde t_2,\vec q^\prime):\rangle \;G(\tilde \omega,\tilde
t_1)G(\tilde \omega^\prime,\tilde t_2),\L{4.6}
\EY
where the Fourier transformation is given by
\BY
&&\delta x_{out,\tilde \omega,\vec q}=\frac{1}{\sqrt {\tilde T_R} }\int_0^{\tilde T_R}d\tilde t \;\delta x^R(\tilde
L,\tilde t,\vec q)\;e^{\ds i\tilde \omega \tilde t}.
\EY

As for the input correlation function $\langle:\delta\hat x_{in, \tilde \omega,\vec q}\;\delta\hat x_{in, \tilde
\omega^\prime,\vec q^\prime}:\rangle$ in the denominator of (\ref{4.4}), one can find it in the explicit form in
Appendix B for two squeezed light sources - the sub-Poissonian phase-locked laser and degenerated optical parametric
oscillator.

\begin{figure}
 \centering
 \includegraphics[height=50mm]{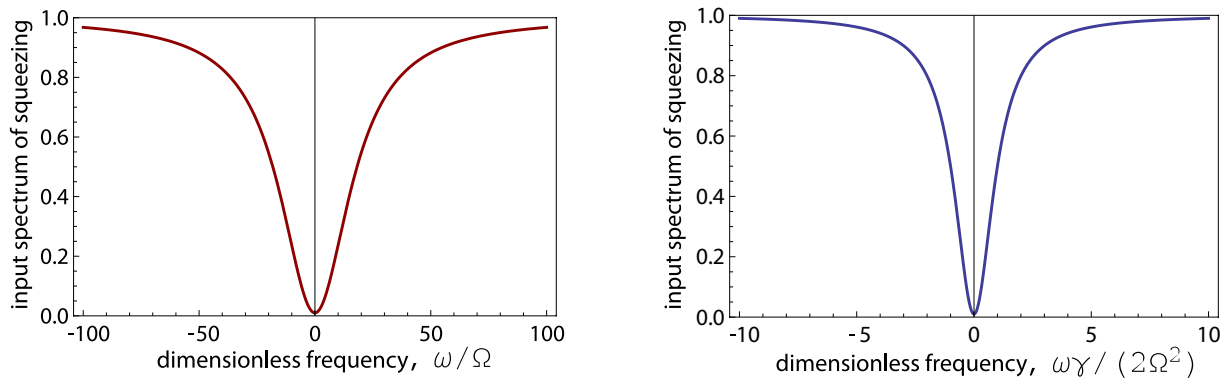}
 \caption{Spectrum of squeezing for the light pulse at the input of QM cell for high-speed (a) and adiabatic (b) QM models. The widths of input
 spectra have been chosen to fulfil the approximate equality $\kappa T \sim 100$. }
 \label{fig2}        
\end{figure}

Before we start to compare the input and output squeezing, it should be noted that the squeezing degree in the input
pulse is already less than in the steady beam \cite{Samb2012}.  This fact is important, because the mechanism of QM
depends essentially on the pulsed nature of the recording light, and we can not neglect by this feature. However, in
\cite{Samb2012} it has been shown that if the condition $\kappa T\gg 1$ for the pulse duration $T$ and the spectral
width of the light $\kappa$ is fulfilled then the pulse squeezing degree does not depend on its duration and equals
to one in stationary radiation. We will assume that this condition is verified. All the necessary details specifying
the input radiation parameters one can find in Appendix B. Graphically, the input pulse spectrum is shown in Fig.
\ref{fig2}.

Bellow we are interested in squeezing of the pulse as a whole that corresponds to the zero frequency component in the
spectrum of squeezing ${\cal S}_{out,\tilde \omega,q}$, i.e., in value ${\cal S}_{out,0,0}$, it characterizes the
maximum achievable degree of squeezing.

It should be noted that the properties of input radiation in two above-described mental experiments are quite
different. In fact, studying the efficiency we are restricted by the analysis of the essentially narrow-band light,
because for given profile of the input it can be characterized by a single time-independent amplitude. That is, one
is discussing the storage for single mode (rectangular in our description). Input-output transformation converts the
profile, resulting in the other field profile at the output, but input and output profiles are connected by unique
integral transformation and the single mode at the input is converted into a single (different from the original)
mode at the output. When we study the degree of squeezing we are dealing with a broadband signal: the input field can
not be characterized by the single amplitude $\hat a_{in}$, but is a function of time. This is the time dependence is
the subject of the study in this experiment. The question of the legitimacy for comparing the results of two such a
different mental experiments, we will discuss below.

\begin{figure}
 \centering
 \includegraphics[height=50mm]{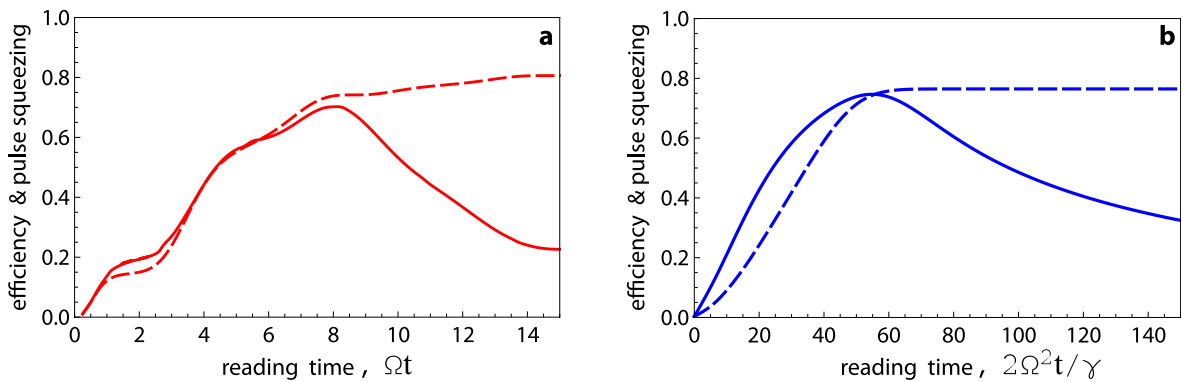}
 \caption{Efficiency (dashed curve) and pulse squeezing (solid curve) as a function of reading time $\tilde T_R$ in two models: high-speed QM (a) and
  adiabatic QM (b) for forward retrieval.}
 \label{fig3}        
\end{figure}

Fig. \ref{fig3} depicts the results of the numerical calculations of the efficiency ${\cal E}$ and pulse squeezing
${\cal S}_{out,0,0}$ (more precisely, $1-{\cal S}_{out,0,0}$) as a function of read-out time $\tilde T_R$ for forward
retrieval in two models: high-speed QM (Fig. \ref{fig3}a) and adiabatic QM (Fig. \ref{fig3}b).

The parameters for numerical calculation have been chosen as follows. In the scheme of high-speed QM for given
dimensionless length of the medium $\tilde L=10$, we found the duration of the input pulse, resulting in the lowest
signal losses in writing (see \cite{HSQM2}), $\tilde T_W=5.5$ meets this requirement. The efficiency in Fig.
\ref{fig3}a is plotted for these parameters. To calculate the pulse squeezing the same values of $\tilde T_W$ and
$\tilde L$ have been chosen, and in addition, we used the input field parameters $\tilde \kappa=100/5.5,\;\;p=1$, to
provide knowingly a good squeezing of the input pulse (below we will demonstrate how to optimize the choice of the
spectral width of the input signal).

For the calculation in the adiabatic limit we have chosen $\tilde T_W=\tilde L=55$, providing approximately the same
efficiency as for the high-speed QM. The value of $\tilde \kappa$ is chosen to satisfy the condition $\tilde \kappa
\tilde T_W\gg 1$. Plot in Fig. \ref{fig3}b corresponds to $\tilde \kappa=100/55$, but it should be noted that the
increase of $\kappa$ has no effect on the behavior of the curves.

First, let us note that the efficiency and pulse squeezing in Fig. \ref{fig3} do not always coincide. The efficiency
increases monotonically and comes to saturation, while the pulse squeezing decreases after a certain time $\tilde
T_R$. Such a behavior of the curves is a clear demonstration of the non-beamsplitter type of interaction in these
models. This result is in line with our knowledge of the role of noises in these models. It is not hard to believe
that we have a high efficiency of QM, but lose the squeezing of radiation due to the contribution of vacuum noise
when the read-out go on too long time.

\begin{figure}
 \centering
 \includegraphics[height=50mm]{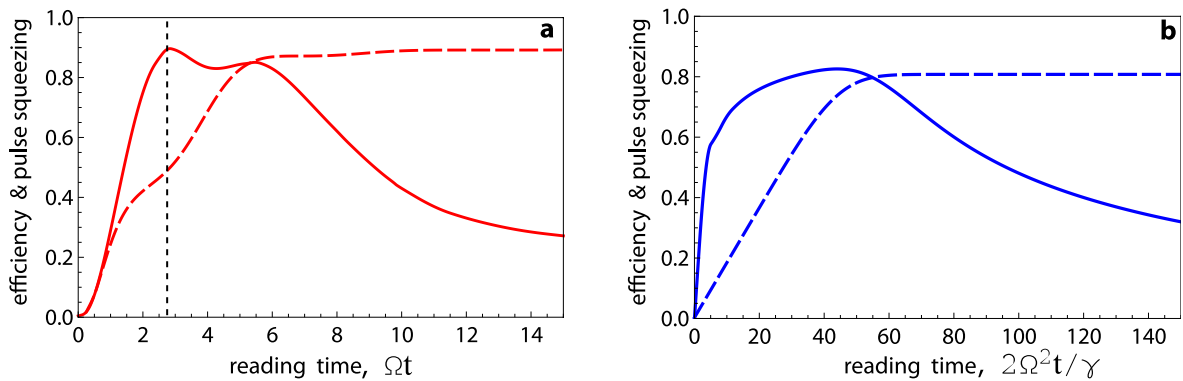}
 \caption{Efficiency (dashed curve) and pulse squeezing (solid curve) as a function of reading time $\tilde T_R$ in two models: high-speed QM (a) and
  adiabatic QM (b) for backward retrieval.}
 \label{fig4}        
\end{figure}

On the other hand, the figure shows the range of values $\tilde T_R$ where pulse squeezing higher than the
efficiency. The same situation takes place for curves in the case of backward retrieval (see Fig.\ref{fig4}).
On the face of it, these curves seem paradoxical: they appeared a range of values $\tilde T_R$ such that under
sufficiently low efficiency (50\% only), we stored a good squeezing in the output. Could it be? A clue to this
paradox lies in the multimode nature of the interaction. In the next section we discuss the mode properties in both
adiabatic and high-speed QM models.

\section{Discussion}
\subsection{Eigenmodes of the memories}
It is not difficult to see that for both considered here models of memory the kernels $G(t,t^\prime)$ in Eqs. (\ref{Ad20}) and (\ref{HS20}), which
couple the input and output signals, are symmetrical with respect to permutation of the arguments $t\leftrightarrow t^\prime$. This means we have a
right to derive the equation for the eigenfunctions $\psi_i(t)$ and eigenvalues $\sqrt{\lambda_i}$ of these kernels in the form
\BY
&&\sqrt{\lambda_i}\psi_i(t)=\int_0^{T_W}dt^\prime\psi_i(t^\prime)G(t,t^\prime).\L{5.1}
\EY
Here the functions $\psi_i(t)$ form the complete ortonormal set:
\BY
&&\int_0^{T_W}dt\psi^\ast_i(t)\psi_j(t)=\delta_{ij},\qquad
\sum_i\psi^\ast_i(t)\psi_i(t^\prime)=\delta(t-t^\prime).\L{5.2}
\EY
The equation (\ref{5.1}) can be solved for each of the four kernels of the integral transformations (\ref{Ad20}) and (\ref{HS20}). In this article we
will analyze the eigenfunctions and eigenvalues only for backward read-out, as this is the case for both models when the "abnormally"\; high pulse
squeezing at the relatively low efficiency appears.

\begin{figure}
 \centering
 \includegraphics[height=50mm]{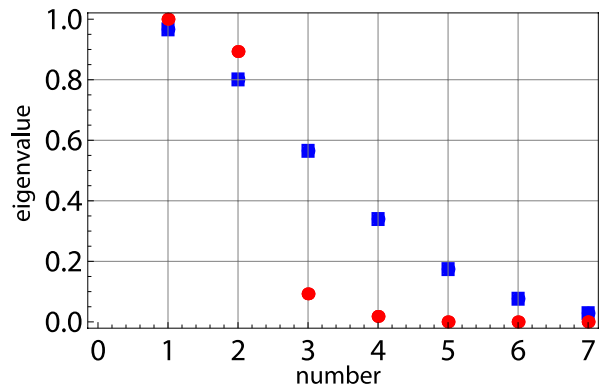}
 \caption{The first seven eigenvalues of the high-speed QM (red, round) and of the adiabatic QM (blue, square).}
 \label{Fig_EV}        
\end{figure}

\begin{figure}
 \centering
 \includegraphics[height=100mm]{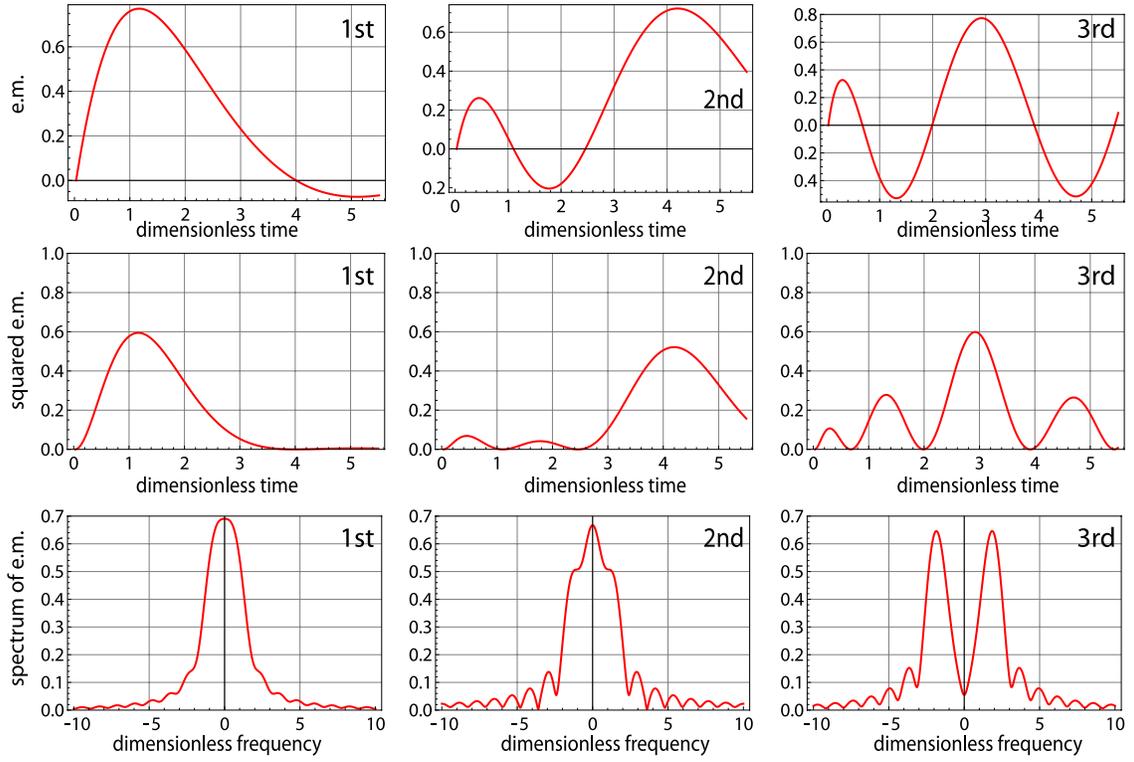}
 \caption{The first three eigenmodes of the integral transformation of the field operator from the input to the output of memory
 cell (upper row),
 squared eigenmodes (middle row) and Fourier spectra of eigenmodes (lower  row) for high-speed QM model. Dimensionless time is
 $\Omega
 t$,
 dimensionless frequency is $\omega/\Omega$.}
 \label{fig5}        
\end{figure}

\begin{figure}
 \centering
 \includegraphics[height=100mm]{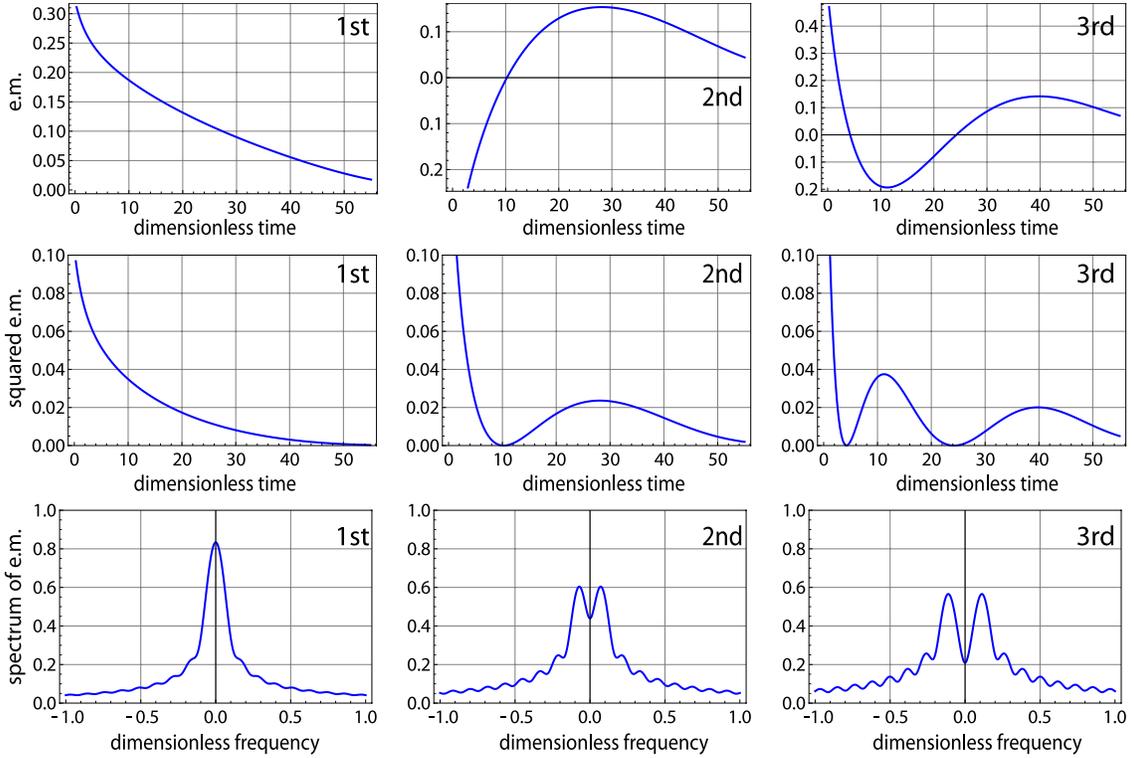}
 \caption{The first three eigenmodes of the integral transformation of the field operator from the input to the output
 of memory cell (upper row),
 squared eigenmodes (middle row) and Fourier spectra of eigenmodes (lower row) for adiabatic QM model. Dimensionless time is $2 \Omega^2 t/\gamma$,
 dimensionless frequency is $\omega\gamma/(2\Omega^2)$.}
 \label{fig6}        
\end{figure}

We will carry out the numerical calculations at the same values of parameters as before (in Fig. \ref{fig3} and \ref{fig4}). First, let us note that
the eigenvalues in both cases decrease rapidly (see Fig. \ref{Fig_EV}).
Especially sharp decrease of eigenvalues is observed for the model of high-speed QM, where, in fact, only the first two eigenvalues differ from zero.
This behavior shows that this scheme is a good filter for the first two eigenmodes. Figures \ref{fig5} and \ref{fig6} shows the first three
eigenmodes (the first row) for the high-speed QM and adiabatic QM models, correspondingly, as well as the squares of these functions (the second
row).

The distinguishing feature of the high-speed QM model is not only the filtering in this scheme of only two modes, but
also the localization of these modes in different intervals of the timeline (the first mode is actually localized in
the range of $ \tilde T_R \in [0,2.75] $, and the second one - in $ \tilde T_R \in [2.75,5.5] $). Draw attention that
the boundaries of intervals coincide with the locations of two peaks in the curve in fig. \ref{fig4}a. For the
adiabatic memory such a mode "separation" is not observed.

Now let us see how squeezing is distributed over the modes, that means we want to follow whether all eigenmodes of the cell are squeezed. For this
aim we plotted the spectra of the first three modes (see the third row in Figs. \ref{fig5} and \ref{fig6}). Recall that the input field has the wide
spectrum of squeezing (see Fig. \ref{fig2}). One can see that all of our spectra of eigenmodes are concentrated in the area of good squeezing, that
means, they are well squeezed. However, not all of them contribute to the squeezing of pulse as whole (squeezing at the zero frequency).

Going back to the curves in Fig. \ref{fig4}a, we see that at time $\tilde T_R = 2.75$ the first mode of the field is
mostly read-out, the second mode is localized on the other time interval, and the rest is filtered out by the cell.
The zero spectral component of the first mode  is high and squeezed well (that is the first mode as whole is squeezed
well). Thus, although the half part of the photons still is not read-out - they will be retrieved from the second
mode, one can obtain a high squeezing in the retrieved pulse.

\subsection{Comparing the actual parameters for two schemes}

It should be noted that the two QM models are very different spectrally. To estimate the typical spectral widths for eigenmodes in both cases, we
should revert from the dimensionless variables to dimensional ones. Recall that the dimensionless procedures for the models are different. For time
we got the relations (\ref{HS29}) and (\ref {Ad4}), respectively:
\BE
\Omega_{HS}\; t \rightarrow \tilde t_{HS},\qquad\frac{2\Omega_{AD}^2}{\gamma}\;t \rightarrow \tilde t_{AD},
\EE
where we now added the below indexes \emph{"HS"}\; and \emph{"AD"}\; to indicate the high-speed or adiabatic model, respectively. Besides, the
requirements imposed on the Rabi frequency, are opposite:
\BE
\gamma\ll\Omega_{HS},\qquad\gamma\gg\Omega_{AD},
\EE
whence it follows $\Omega_{AD}\ll\Omega_{HS}$, and the ratio of this frequencies has to be at least 2 orders. Based
on these inequalities and noting that the ratio of dimensionless spectral widths in Figs. \ref{fig5} and \ref{fig6}
is $\Delta\omega_{HS}/\Delta\omega_{AD} \approx 10$, we find that the width of eigenmodes in the adiabatic model is
at least 4 orders of magnitude smaller than the corresponding width in the high-speed model. At the same time, one
can see that although the dimensionless values of $L$ are different, we can assume that both calculations were
carried out at the same optical depth  $d=2g^2NL/\gamma \sim 55$ (supposing the relation $\Omega_{HS}/\gamma\approx
10$ to be fulfilled). Let us note that the relation between the parameters typical for the high-speed QM, can be
easily reached in experiments with resonant $\Lambda$-atoms, especially for cold atomic ensemble, than for the
adiabatic condition.

Figures \ref{fig5} and \ref{fig6} shows that the spectral bandwidths of memories are much less than the chosen
spectral width of the input signal. We can say that this choice was not optimal, and it is necessary to match the
input signal to the memory capacity. In this sense, the model of high-speed QM looks more attractive because it
provides a much wider spectral bandwidth of the signal as mentioned above.

\begin{figure}
 \centering
 \includegraphics[height=50mm]{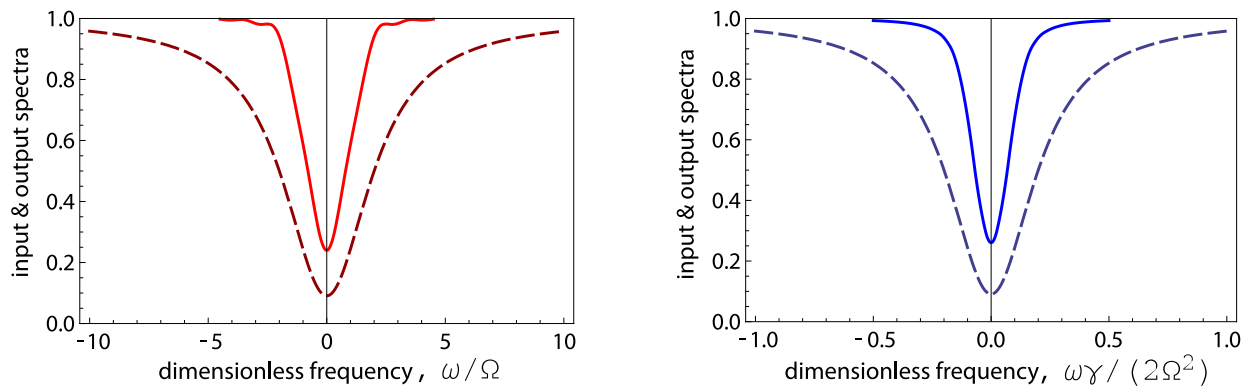}
 \caption{Input and retrieval spectra of squeezing for high-speed (a) and adiabatic (b) QM models.}
 \label{fig8}        
\end{figure}

The decrease in the value of the spectral width $\kappa$, while maintaining the duration of the input pulse results
in a loss of input pulse squeezing (see Fig. \ref{fig8}). However, it is seen that for $\tilde \kappa \approx 2$ for
the high-speed QM and for $\tilde \kappa \approx 0.2$ for the adiabatic QM these losses are relatively small (about
10\%), and the ratio $\tilde \kappa \tilde T \gg 1$ is still satisfied.

In Fig. \ref{fig8} the input and retrieval spectra of squeezing are plotted for the case when spectral bandwidth of
squeezing of the input signal is matched to the memory bandwidth.

Since considered memory cells are the mode filters, i.e. they are sensitive to only a few the first modes of input signal, then the presence of other
modes at the input does not affect to the pulse squeezing. From the Fig. \ref{fig5} one can see that the zero spectral component is large enough only
in the spectra of the first and second eigenmodes. At the same time, the square input pulse is well approximated by the superposition of the first
two modes. That is why the comparison of the two mental experiments (estimation of the efficiency and of the storage of squeezing) with the different
input signals seems to us justified. For example, in the scheme of high-speed QM if knowing the properties of our cell as a mode filter, we form an
input signal as a superposition of the first two modes, then the curves in Fig. \ref{fig4} remain almost unchanged.

\section{Is the efficiency of exhaustive characteristic of quantum memory?}

In Raman memory models based on three-level atoms, two quantum sub-systems (the signal field and atomic coherence) are interacted with each other.
The writing and read-out processes are formally realized by means of the interaction Hamiltonian of the kind of $g(\hat a^\dag\hat b+\hat a\hat
b^\dag)$. This Hamiltonian ensures the beamsplitter-like interaction for the thin atomic layer and as a result one can derive the input-output
equation (\ref{Eff1}) and conclude that squeezing and efficiency are coupled by the formula (\ref{Eff6}). This means that the efficiency turns out to
be sufficient to assess the quality of memory \cite{Gorshkov_cavity,Gorshkov_free}.

In this article we consider more complicated situation, which takes place in the cases of adiabatical and high-speed memories in the strongly
resonant limit.  In each of these models there are three quantum sub-systems: the signal field and two atomic coherences for the high-speed memory
and the signal field, atomic coherence, and the vacuum reservoir, forming the Langevine source, for the adiabatic memory. The presence of more than
two active subsystems makes impossible direct interpretation of interaction as a beamsplitter-like one. However maybe we still can apply such an
interpretation in some sense? Let's consider again the equation (\ref{Ad22}) coupling the signal field at the output of the memory cell after the
read-out with the signal at the front of the cell before the writing process. Let's rewrite it in the form of the non-averaged over vacuum
sub-systems
\BY
&&\hat a_{out }(\tilde t,\vec q)=\int_0^{\tilde T_W} G(\tilde t,\tilde t^\prime)\hat a_{in}(\tilde T_W-\tilde t^\prime,\vec q)d\tilde t^\prime +\hat
F_{vac }(\tilde t,\vec q).\L{6.1}
\EY
Remind the indexes "in" \;and "out" \;mean $\tilde z=0$ and $\tilde z=\tilde L$, respectively. The second term on the right $\hat F_{vac }$
concentrates contributions of all sub-systems, which initially were in the vacuum state.
It is not difficult to see that the input-output amplitudes obey the correct commutation relations
\BY
&&\[\hat a_{out }(\tilde t_1,\vec q_1),\hat a^\dag_{out }(\tilde t_2,\vec q_2)\]=\[\hat a_{in}(\tilde t_1,\vec q_1 ),\hat a^\dag_{in}(\tilde t_2,\vec
q_2)\]=\delta(\tilde t_1-\tilde t_2)\delta^2(\vec q_1-\vec q_2).\L{ }
\EY
provided that
\BY
&&\[\hat F_{vac }(\tilde t_1,\vec q_1),\hat F_{vac }(\tilde t_2,\vec q_1)\]=\(\delta(\tilde t_1-\tilde t_2)-\int_0^{\tilde T_W} G(\tilde t_1,\tilde t
)G(\tilde t_2,\tilde t)d\tilde t\) \delta^2(\vec q_1-\vec q_2) \L{6.3}.
\EY
As mentioned above the kernel $G(\tilde t,\tilde t^\prime)$ is symmetrical and  we have a right to derive Eq. (\ref{5.1}) for its eigenfunctions,
which obey the requirements (\ref{5.2}).

Let us decompose input and output amplitudes in the form
\BY
&&\hat a_{in}(\tilde T_w-\tilde t,\vec q)=\sum_i\hat e_{in;i,\vec q}\;\psi_i (\tilde t),\qquad\hat a_{out}(\tilde t,\vec q)=\sum_i\hat e_{out;i,\vec
q}\;\psi_i (\tilde t),\L{6.6}
\EY
where
\BY
&& \[\hat e_{in;i,\vec q},\hat e^{\dag}_{in;i^\prime,\vec q^\prime}\]=\[\hat e_{out;i,\vec q},\hat e^{\dag}_{out;i^\prime,\vec
q^\prime}\]=\delta_{ii^\prime}\delta^2(\vec q-\vec q^\prime).
\EY
The similar decomposition can be used for function $\hat F_{vac}(\tilde t,\vec q)$:
\BY
&&\hat F_{vac}(\tilde t,\vec q)=\sum_i\hat f_{vac;i,\vec q}\;\psi_i (\tilde t),\quad \mbox{where}\quad \hat f_{vac;i,\vec q}=\int_0^{\tilde T_W}
d\tilde t\psi_i (\tilde t)\hat F_{vac}(\tilde t,\vec q) . \L{6.7}
\EY
Taking into account the equation (\ref{6.3}) one can obtain
\BY
&&\[\hat f_{vac;i,\vec q},\hat f^{\dag}_{vac;i^\prime,\vec q^\prime}\]=(1-\lambda_i)\delta_{ii^\prime}\delta^2(\vec q-\vec q^\prime) \L{}
\EY
or making normalization
\BY
&&\[\hat e_{vac;i,\vec q},\hat e^{\dag}_{vac;i^\prime,\vec q^\prime}\]=\delta_{ii^\prime}\delta^2(\vec q-\vec q^\prime),\quad\mbox{where}\quad\hat
f_{vac;i,\vec q}=\sqrt{1-\lambda_i}\;\hat e_{vac;i,\vec q} . \L{6.8}
\EY
Substituting (\ref{6.6}), (\ref{6.7}) and (\ref{6.8}) into (\ref{6.1}) one can obtain for each of modes
\BY
&&\hat e_{out;i,\vec q}=\sqrt{\lambda_i} \;\hat e_{in;i,\vec q}-\sqrt{1- \lambda_i}\;\hat e_{vac;i,\vec q}.\L{6.11}
\EY
Thus, in this sense, we can consider our model as a beamsplitter-like ones. However such a classification associated with choice of the input signal
as one (and only one) of the memory eigenfunction.

The question arises whether we can correctly determine the efficiency for an arbitrary input signal and whether it will be as comprehensive
characteristic of the memory as the efficiency associated with the transfer of eigenmode. If we consider the arbitrary input signal as a single mode
from some complete orthonormal set of functions $ \hat A_{in}(T_W-t) = h_1(t)\hat E_1$, then the efficiency can be expressed analytically. In this
case, one can obtain from the formula (\ref{4.2})
\BY
&& {\cal E}= \int_0^{\tilde T_R} d\tilde t \(\int_0^{\tilde T_W} d\tilde t^\prime h_1 (\tilde t^\prime)G(\tilde t,\tilde t^\prime)\)^2.\L{E4}
\EY
To assess how squeezing is changed under the full memory process, we can use ratio (\ref{4.4}) putting there $\omega=0$ and $\vec q=0$
\BY
&& {\cal E}_{squuzing}=\frac{\langle: \delta\hat x_{out,\omega=0,\vec q=0}\delta\hat x_{out,-\omega=0,-\vec q=0} :\rangle}{\langle: \delta\hat
x_{in,\omega=0,\vec q=0}\delta\hat x_{in,-\omega=0,-\vec q=0} :\rangle}.
\EY
According (\ref{4.6}) the expression in numerator is given by
\BY
&& \langle:\delta\hat x_{out, \omega=0,\vec q=0}\;\delta\hat x_{out, -\omega=0,-\vec q=0}:\rangle=\nn\\
&&=\int_0^{\tilde T_W} d\tilde t_1 \int_0^{\tilde T_W} d\tilde t_2\langle:\delta\hat x_{in}(\tilde T_W-\tilde t_1,\vec q=0)\;\delta\hat x_{in}(\tilde
T_W-\tilde t_2,-\vec q=0):\rangle \;G(\omega=0,\tilde t_1)G(-\omega=0,\tilde t_2).
\EY
Now taking into account that
\BY
&&  \langle:\delta\hat x_{in, \omega=0,\vec q=0}\;\delta\hat x_{in, -\omega=0,-\vec q=0}:\rangle=\frac{1}{2\pi}\int_0^{\tilde T_W} d\tilde t_1
\int_0^{\tilde T_W} d\tilde t_2\langle:\delta\hat x_{in}(\tilde t_1,\vec q=0)\;\delta\hat x_{in}(\tilde t_2,-\vec q=0):\rangle .
\EY
and $\hat A_{in}(\tilde t)=h(\tilde t)\hat E_1$ one can obtain
\BY
&& {\cal E}_{squeezing}=\left[{\int_0^{\tilde T_R} d\tilde t \int_0^{\tilde T_W} d\tilde t^\prime h_1 (\tilde t^\prime) G(\tilde t,\tilde
t^\prime)/\int_0^{\tilde T_W} h_1 (\tilde t) d\tilde t}\right]^2.\L{E6}
\EY
It is clear that the expressions (\ref{E4}) and (\ref{E6}) coincide only if the profile $ h_1(\tilde t)$ is an eigenfunction of the kernel $G(\tilde
t, \tilde t^\prime)$. Thus, the efficiency determined for a single non-eigenmode provides an answer to the question about the ability of the memory
to store the total number of photons, but does not characterize the preservation of other quantum properties of light, such as squeezing.

Beamsplitter relation can be generalized in the form \cite{Sokolov}:
\BY
&&\hat r=\sqrt{\lambda} \;\hat s - \sqrt{1- \lambda}\;\hat e_{vac},\L{6.15}
\EY
where operator $\hat s$ is a quantum amplitude of single input mode of the signal, and $\hat r$ is an amplitude of different mode of output signal,
which can be defined as a projection of input mode on the full set of the eigenmodes $\psi_i (t)$ of the kernel of integral transformation
(\ref{6.1}). It is easy to see that this equation turns into Eq. (\ref{6.11}) when the input field is chosen in the form of one of the modes $\psi_i
(t)$. Eq. (\ref{6.15}) can be used for comparison of the degree of squeezing in the input and output modes. However it does not provide us with any
information about the ratio of squeezing on given (for example, zero) frequency that is the most important parameter for multimode squeezing. We
should note that Eq. (\ref{6.15}) is valid only when writing time is equal to reading time. Moreover, it links two single modes - one mode at the
input and one mode at the output - and is inapplicable for multimode stochastic fields considered here. Such a generalization can be a subject for
further study.

\section{Conclusion}

Addressing once again to the comparison of two memory models presented in this paper, we note that different factors lead to a vanishing of the role
of the Langevin noises in this cases. In high-speed QM model the noises can be excluded from consideration in very beginning, when we construct
equations, since on the typical interaction time the spontaneous decay of the excited state is negligible, and hence the noise associated with this
decay is also small. The situation is different in adiabatic QM model. Here, the rate of the spontaneous relaxation is large that yields to the
presence of Langevin noise sources, which associated with population decay of the excited state as well as with decay of the coherences on the
transitions coupled with the excited state. However, due to the initial assumption that the population of level $ | 3 \rangle $ is small (and almost
all atoms remain on the level $ | 1 \rangle $) the additional noise terms turn out to be small.

In this study, we demonstrated that both memory models considered here allow us to achieve high efficiency for writing and restore of the light
pulse, and are suitable for storage of the squeezed states.

We have shown that the quantum efficiency of whole writing-reading cycle for a pulse with arbitrary time profile is not a universal characteristic of
the memory, through which the other properties of the memory (for example, the ability to store a squeezing) can be expressed. Only the efficiency
defined in relation to a single eigenmode of the memory is such a comprehensive parameter. Thus for a pulse with an arbitrary shape it is not enough
to provide the high efficiency of the memory to conclude about preserving the quantum state of light.

We have analyzed the eigenmodes of the two memory models discussed here and demonstrated interesting feature connected with the ability to manipulate
separately by two first (high-efficient) eigenmodes in the high-speed QM scheme.

Since the quantum memory operates as a mode filter, its important characteristic is a spectral bandwidth. Estimating the spectral bandwidth of the
adiabatic and high-speed QM schemes, we have shown that it is significantly different for these schemes. The spectral bandwidth of the high-speed QM
is at least 4 orders of magnitude higher than of the adiabatic QM. This demonstrate the benefits of the high-speed memory for storage of wideband
signals.

\section{Acknowledgements} We thank Elisabeth Giacobino and Ivan Sokolov for productive discussions. The reported study was supported by RFBR
(grants No. 12-02-00181a and 13-02-00254a) and by the ERA.Net RUS Project NANOQUINT.

\appendix

\section{General solutions for high-speed memory model}\L{A}
The general solutions of the Eqs. (\ref{HS24})-(\ref{HS26}) under the arbitrary initial and boundary conditions can
be written over the non-canonical amplitudes, introduced analogously to our consideration in Section \ref{NCA_Ad}:
\BY
&&\hat A^W( \tilde z, \tilde t;\vec q)=\int_0^{  \tilde t}d  \tilde t^\prime \hat a_{in}( \tilde t-  \tilde t^\prime;\vec q)
\;G_{aa}(\tilde z,  \tilde t^\prime),\L{A1}\\
&&\hat B^W(\tilde z,  \tilde t;\vec q)=-\frac{1}{p}\;\int_0^{  \tilde t}d  \tilde t^\prime \hat a_{in}
(\tilde  t-  \tilde t^\prime;\vec q)\;G_{ab}(\tilde z,\tilde t^\prime),\L{A2} \\
&&\hat C^W(\tilde  z, \tilde t;\vec q)=\frac{1}{p}\int_0^{  \tilde t}d  \tilde t^\prime \hat a_{in}( \tilde t- \tilde
t^\prime;\vec q)\;G_{ac}( \tilde t^\prime, \tilde z),\L{A3}
\EY
where $p={\Omega}/{(g\sqrt N)}$.
\BY
&&\hat A^R( \tilde z,\tilde  t;\vec q)=-\frac{p}{2}\int_0^{  \tilde z}d  \tilde z^\prime \hat B^R(\tilde z-\tilde
z^\prime, 0;\vec q)\;G_{ba}(\tilde z^\prime, \tilde t ),\quad\mbox{where}\quad\hat
B^R(\tilde z,  0;\vec q)=\hat B^W(\tilde z ,  \tilde T_W;\vec q),\L{A4}\\
&&\hat B^R(\tilde z,  \tilde t;\vec q)=\frac{1}{2}\;\int_0^{  \tilde z}d  \tilde z^\prime \hat B^R(\tilde z-\tilde
z^\prime, 0;\vec
q)\;G_{bb}(\tilde z^\prime,\tilde t) ,\L{A5}\\
&&\hat C^R( \tilde z, \tilde t;\vec q)=\frac{1}{2}\;\int_0^{  \tilde z}d  \tilde z^\prime \hat B^R(\tilde z-\tilde
z^\prime,0;\vec q)\;G_{bc}(\tilde \tilde z^\prime, t ),\L{A6}
\EY
where upper indexes "W"\; and "R"\; indicate the writing and read-out stage, respectively.

Here kernels $G_{ik}(\tilde z^\prime,  \tilde t)$ are bilinear combinations of the expressions depending on the n-th
Bessel functions of the first kind denoted by $J_{n}$
\BY
&&g_{aa}(\tilde z,  \tilde t)=\delta(  \tilde t)-e^{\ds -i  \tilde t}\sqrt{{  \tilde z}/{(4  \tilde t)}}\;
J_1\(\sqrt{  \tilde t  \tilde z}\)\Theta(  \tilde t),\L{A7}\\
&&g_{bb}(\tilde z,  \tilde t)=e^{\ds-i  \tilde t}\;\sqrt{{4  \tilde t}/{   \tilde z}}\;J_1\(\sqrt{  \tilde t  \tilde z}\)
\Theta(  \tilde t),\L{A8}\\
&&g_{ab}(\tilde z,  \tilde t)=e^{\ds-i  \tilde t}\; J_0\(\sqrt{  \tilde t  \tilde z}\)\Theta(  \tilde t).\L{A9}
\EY
The kernels in Eqs.~(\ref{A1})-(\ref{A6}) read
\BY
&&G_{aa}(\tilde z, \tilde  t)=\int_0^{\tilde t}d\tilde t^\prime g_{aa}(\tilde z,\tilde t-\tilde t^\prime,)
g_{aa}^\ast(\tilde z,\tilde t^\prime),\L{A10}\\
&&G_{ab}(\tilde z,  \tilde t)=\int_0^{\tilde t}d\tilde t^\prime g_{ab}(\tilde z,\tilde t-\tilde t^\prime,)
g_{ab}^\ast(\tilde z,\tilde t^\prime),\L{A11}\\
&&G_{ac}(\tilde z,  \tilde t)=\frac{1}{2}\int_0^{\tilde t}dt^\prime g_{aa}(\tilde z,\tilde t-\tilde t^\prime,)
g_{ab}^\ast(\tilde z,\tilde t^\prime)+c.c.,\L{A12}\\
&&G_{ba}(\tilde z,  \tilde t)=G_{ab}(\tilde z,  \tilde t),\L{A13}\\
&&G_{bb}(\tilde z,  \tilde t)=2\;\delta(  \tilde z)\cos \tilde t+
\int_0^{\tilde t}d\tilde t^\prime g_{bb}(\tilde z,\tilde t-\tilde t^\prime,)g_{bb}^\ast(\tilde z,\tilde t^\prime),\L{A14}\\
&&G_{bc}(\tilde z,  \tilde t)= \delta(  \tilde z)\sin \tilde t-\frac{1}{2}\int_0^{\tilde t}d\tilde t^\prime
g_{aa}(\tilde z,\tilde t-\tilde t^\prime,)g_{ab}^\ast(\tilde z,\tilde t^\prime)+c.c.\L{A15}
\EY
\section{Squeezed pulse radiation from sub-Poissonian laser and DOPO}

\subsection{Stationary radiation}

According to \cite{SPL}, the mean-square intracavity quadrature fluctuations of a single-mode laser are determined as
\BY
&&(:\delta  x_\omega^2:)=-\frac{p(1-\mu)}{4}\frac{\kappa}{\kappa^2(1-\mu/2)^2+\omega^2},\qquad(:\delta
y_\omega^2:)=\frac{1-\mu}{2}\frac{\kappa}{\kappa^2\mu^2/4 +\omega^2},\qquad\mu=\sqrt{\frac{n_0}{n}}\ll1.\L{16}
\EY
Here, the parameter $0 < p < 1$ reflects the degree of ordering of the upper laser level excitation: $p = 0$ for a completely random Poissonian
pumping, and $p = 1$ for a completely regular one. Here, $\kappa$ is the spectral width of the laser mode, $n$ is the average number of photons in
the oscillation mode in the steady-state regime of oscillation and $n_0$ is the average number of photons appearing in an empty cavity under the
influence of an external locking field. This locking is necessary for suppression of phase diffusion in the oscillation, which prohibits efficient
quadrature squeezing. Parameter $\mu$ determining laser locking to an external field is assumed to be small compared to unity. However, despite small
value of $\mu$, ignoring this quantity would violate the Heisenberg uncertainty relation at small frequencies. Therefore, this quantity should be
handled carefully.

Similarly, we can write the corresponding expressions for the case of a degenerate optical parametric generation \cite{DOPO1}:
\BY
&&(:\delta  x_\omega^2:)=\frac{1}{4}\;\frac{\kappa s}{\kappa^2/4(1-s)^2+\omega^2},\qquad(:\delta y_\omega^2:)=-\frac{1}{4}\;\frac{\kappa
s}{\kappa^2/4(1+s)^2+\omega^2},\qquad s=\sqrt{{n_p}/{n_0}}<1.\L{17}
\EY
Here, $n_p$ is the average steady-state number of photons stored in the pump mode, $n_0$ is the threshold number of photons in the pump mode, and
$\kappa$ is the spectral width of the signal mode. Here, we analyze the operation of a DOPO in the below-threshold regime and the parameter $s$
characterizes the degree of approaching the generation threshold.

\subsection{Spectrum of squeezing in isolated pulse of light}

Applying inverse Fourier transformation to (\ref{16}), we have
\BY
&&\langle:\delta x(t)\;\delta x(t^\prime):\rangle=-\frac{p}{8}\;\frac{1-\mu}{1-\mu/2}\;e^{\ds
-\kappa(1-\mu/2)|t-t^\prime|},\L{18}\\
&&\langle:\delta y(t)\;\delta y(t^\prime):\rangle=\frac{1-\mu}{\mu}\;e^{\ds -\kappa\mu/2|t-t^\prime|}.\L{19}
\EY
When describing a pulse of light in the quantum theory, it should be remembered that the field outside of the pulse is in a vacuum state, which can
be explicitly expressed by writing the extracavity field amplitude in the form
\BY
&&\hat A(t)\to\Theta^T(t)\hat A(t)+\(1-\Theta^T(t)\)\hat A_{vac }(t).\L{20}
\EY
Here, the index \emph{"vac"}\; in front of the Heisenberg operator means that the operator must be averaged over the vacuum state when calculating
measurable quantities. The function $\Theta^T(t)$ is zero outside of the $0 < t < T$ interval and is equal to unity within this interval.

To characterize squeezing, along with the field quadratures for the intracavity field $\hat x(t)$ and $\hat y(t)$, the quadratures for the field
leaving the cavity for free-space propagation
\BY
&&\hat X(t)=\frac{1}{2}\(\hat A^\dag(t)+\hat A(t)\),\qquad\hat Y(t)=\frac{i}{2}\(\hat A^\dag(t)-\hat A(t)\).
\EY
are introduced. Quadrature squeezing for stationary generation is determined based on the equalities relating the correlator of the quadrature
operators with the corresponding average of the normal-ordered operators:
\BY
&&\langle \hat X(t) \hat X(t^{\prime}) \rangle = \frac{1}{4}\delta(t-t^{\prime})+ \kappa\langle : \hat x(t) \hat x(t^{\prime}) :\rangle.\L{7}
\EY
\BY
&&\langle \hat Y(t) \hat Y(t^{\prime}) \rangle = \frac{1}{4}\delta(t-t^{\prime})+ \kappa\langle : \hat y(t) \hat y(t^{\prime}) :\rangle.\L{8}
\EY
For the isolated pulse equations (\ref{7}) and (\ref{8}) can be written in the form
\BY
&&\langle \Theta^T(t)\delta\hat X(t) \;\Theta^T(t^{\prime})\delta\hat X(t^{\prime}) \rangle =
\frac{1}{4}\Theta^T(t)\Theta^T(t^{\prime})\;\delta(t-t^{\prime})+ \kappa\langle : \Theta^T(t)\delta\hat x(t)
\;\Theta^T(t^{\prime})\delta\hat x(t^{\prime}) :\rangle,\L{21}\\
&&\langle \Theta^T(t)\delta\hat Y(t) \;\Theta^T(t^{\prime})\delta\hat Y(t^{\prime}) \rangle =
\frac{1}{4}\Theta^T(t)\Theta^T(t^{\prime})\;\delta(t-t^{\prime})+ \kappa\langle : \Theta^T(t)\delta\hat y(t) \;\Theta^T(t^{\prime})\delta\hat
y(t^{\prime}) :\rangle.\L{22}
\EY
Let us define the Fourier transformation for function $F(t)$ in limited time interval $[0, T]$ as
\BY
&&F^T_\omega=\frac{1}{\sqrt T }\int_0^T dt F(t)e^{i\omega t}\;.\L{23}
\EY
Application of this transformation to (\ref{21}) and (\ref{22}) yields
\BY
 && \langle\delta\hat X^T_{\omega} \;\delta\hat X^T_{\omega^{\prime}} \rangle =  \frac{1}{4}
\delta^T(\omega+ \omega^{\prime})+\kappa\langle:\delta\hat x^T_{\omega} \;\delta\hat x^T_{\omega^{\prime}}:
\rangle,\L{24}\\
 && \langle\delta\hat Y^T_{\omega} \;\delta\hat Y^T_{\omega^{\prime}} \rangle =  \frac{1}{4}
\delta^T(\omega+ \omega^{\prime})+\kappa\langle:\delta\hat y^T_{\omega} \;\delta\hat y^T_{\omega^{\prime}}: \rangle,\L{25}
\EY
where we introduced the notation
\BY
 &&\delta^T(\omega+ \omega^{\prime})=
 \frac {\sin(\omega+ \omega^{\prime})T/2}{(\omega+ \omega^{\prime})T/2}\;
 e^{\ds { i (\omega + \omega^{\prime}) T}/{2}}\;.\L{26}
\EY
The second terms in (\ref{24})-(\ref{25}), which are related to calculation of the normally ordered average, are different for different sources. In
the case of single-mode phase-locked sub-Poissonian laser, we readily find that
\BY
&& \langle:  \delta\hat x^T_{\omega}\; \delta\hat  x^T_{\omega^{\prime}} :\rangle =\frac{p}{8}\frac{1-\mu}{1-\mu/2}\[- \(
\frac{1}{\kappa_x+i\omega^{\prime}} + \frac{1}{\kappa_x+i\omega} \)\delta^T(\omega+\omega^{\prime}) + \right.\nn\\
&&\left.+\frac{1}{T}   \frac{1}{(\kappa_x-i\omega)(\kappa_x+i\omega^{\prime})} \( {1-e^{\ds -\kappa_x T+ i\omega T}}{} \)
+\frac{1}{T}  \frac{1}{(\kappa_x-i\omega^{\prime})(\kappa_x+i\omega)}  \( {1- e^{\ds -\kappa_x T+ i\omega^{\prime} T}} \)\],\L{27}
\EY
where
$$
\kappa_x=\kappa(1-\mu/2)
$$
Using definition (\ref{Eff4}) and substituting it into (\ref{24}) and (\ref{27}), we can express the squeezing parameter for the case of a
sub-Poissonian laser in the form
\BY
 && e^{\ds -r_{in}(\omega)} =
 1- \frac{p\kappa^2(1-\mu)}{\kappa_x^2 + \omega^2}   +
\frac{p\kappa^2(1-\mu)}{2 \kappa_x T} \[\frac{1}{(\kappa_x - i \omega)^2}\(1 - e^{\ds -\kappa_x T +i \omega T}\)+ \mbox{c.c.}
\].\L{28}
\EY

For parametric oscillation, the most interesting feature is the phase quadrature. Formally acting similar to the laser case, i.e., limiting analysis
to time interval $0 < t < T$ and applying Fourier transformation, we have
\BY
 && \langle  \delta\hat Y^T_{\omega}\; \delta\hat  Y^T_{\omega^{\prime}} \rangle  =
  \frac{1}{4}\[1-\frac{\kappa s}{1+s} \(
\frac{1}{\kappa/2(1+s)+i\omega^{\prime}} + \frac{1}{\kappa/2(1+s)+i\omega} \)\]\delta^T(\omega+\omega^{\prime})+\nn\\
&& +\frac{\kappa}{4 T}\frac{s}{1+s} \frac{1}{(\kappa/2(1+s)-i\omega)(\kappa/2(1+s)+i\omega^{\prime})} \( {1-e^{\ds -\kappa/2(1+s) T+
i\omega T}} \)- \nn\\
&& +\frac{\kappa}{4 T}\frac{s}{1+s} \frac{1}{(\kappa/2(1+s)-i\omega^{\prime})(\kappa/2(1+s)+i\omega)} \( {1-e^{\ds -\kappa/2(1+s) T+ i\omega^{\prime}
T}} \).\L{35}
 \EY
The obtained expression allows estimating the degree of squeezing of a pulse of light of arbitrary duration $T$ at arbitrary frequency.


\begin{thebibliography}{100}
%
\bibitem{repeaters1} L. M. Duan, M. D. Lukin, J. I. Cirac, and P. Zoller,
Long-distance quantum communication with atomic ensembles and linear optics, Nature, 414(6862), 413-418 (2001)
%
\bibitem{repeaters2} Sangouard N, Simon C, Zhao B et al., Phys. Rev. A, 77(6), 062301 (2008)
%
\bibitem{repeaters3} Nicolas Sangouard, Christoph Simon, Hugues de Riedmatten, and Nicolas Gisin,
Quantum repeaters based on atomic ensembles and linear optics, Rev. Mod. Phys., 83, 33-80  (2011)
%
\bibitem{comp1}  E. Knill, R. Laflamme, and G. J. Milburn, A scheme for efficient quantum computation with
linear optics, Nature, 409(6816), 46-52 (2001)
%
\bibitem{comp2}  Pieter Kok, W. J. Munro, Kae Nemoto, T. C. Ralph, Jonathan P. Dowling, and G. J. Milburn,
Linear optical quantum computing with photonic qubits, Rev. Mod. Phys., 79(1), 135 (2007)
%
\bibitem{appl}  F. Bussi\`{e}res, N. Sangouard, M Afzelius, H. de Riedmatten, C. Simon, and W. Tittel, Prospective
applications of optical quantum memories, arXiv:1306.6904 (2013)
%
\bibitem{review1} Klemens Hammerer, Anders S. Sorensen, and Eugene S. Polzik,
Rev. Mod. Phys., 82, 1041-1093 (2010)
%
\bibitem{review2}  Alexander I. Lvovsky, Barry C. Sanders, and Wolfgang Tittel, Optical quantum memory,
Nature Photonics, 3(12), 706-714 (2009)
%
\bibitem{ex1} Chien Liu, Zachary Dutton, Cyrus H. Behroozi, and Lene V. Hau,
Observation of coherent optical information storage in an atomic medium using halted light pulses, Nature, 409(6819), 490-493 (2001)
%
\bibitem{ex2} Brian Julsgaard, Jacob Sherson, J. Ignacio Cirac, Jaromir Fiurasek, and Eugene S. Polzik,
Experimental demonstration of quantum memory for light, Nature , 432(7016), 482-486 (2004)
%
\bibitem{ex3} M. D. Eisaman, A. Andr\'{e}, F. Massou, M. Fleischhauer, A. S. Zibrov, and M. D. Lukin,
Electromagnetically induced transparency with tunable single-photon pulses, Nature , 438(7069), 837-841 (2005)
%
\bibitem{ex4} K. S. Choi, H. Deng, J. Laurat, and H. J. Kimble,
Mapping photonic entanglement into and out of a quantum memory, Nature , 452(7183), 67-71 (2008)
%
\bibitem{ex5} Bo Zhao, Yu-Ao Chen, Xiao-Hui Bao, Thorsten Strassel, Chih-Sung Chuu, Xian-Min Jin,
Jorg Schmiedmayer, Zhen-Sheng Yuan, Shuai Chen, and Jian-Wei Pan, A millisecond quantum memory for scalable quantum networks, Nature Physics, 5(2),
95-99 (2008)
%
\bibitem{ex6} K. F. Reim, J. Nunn, V. O. Lorenz, B. J. Sussman, K. C. Lee, N. K. Langford, D. Jaksch, and I. A. Walmsley,
Towards high-speed optical quantum memories, Nature Photonics, 4, 218-221 (2010)
%
\bibitem{ex7} U. Schnorrberger, J. D. Thompson, S. Trotzky, R. Pugatch, N. Davidson, S. Kuhr, and I. Bloch,
Electromagnetically induced transparency and light storage in an at omic mott insulator, Phys. Rev. Lett., 103, 033003 (2009)
%
\bibitem{ex8} B. Kraus, W. Tittel, N. Gisin, M. Nilsson, S. Kr\"{o}ll, and J. I. Cirac,
Quantum memory for nonstationary light fields based on controlled reversible inhomogene ous broadening, Phys. Rev. A, 73, 020302 (2006)
%
\bibitem{Nunn2013} J. Nunn, N. K. Langford, W. S. Kolthammer, T. F. M. Champion, M. R. Sprague, P. S. Michelberger,
X.-M. Jin, D. G. England, I. A. Walmsley, Phys. Rev. Lett. 110(13), 133601 (2013)
%
\bibitem{Nunn2008} Nunn J, Reim K, Lee KC et al., Phys. Rev. Lett., 101(26), 260502 (2008)
%
\bibitem{Simon} Simon C, de Riedmatten H, Afzelius M et al., Phys. Rev. Lett., 98, 190503 (2007)
%
\bibitem{high-bandwidth} Valentina Caprara Vivoli, Nicolas Sangouard, Mikael Afzelius, Nicolas Gisin, New J. Phys., 15, 095012 (2013)
%
\bibitem{Sorensen} Ivan Iakoupov, Anders S. Sorensen, New J. Phys. 15, 085012 (2013)
%
\bibitem{Tittel} Neil Sinclair, Erhan Saglamyurek, Hassan Mallahzadeh, Joshua A. Slater, Mathew George, Raimund Ricken, Morgan P. Hedges, Daniel Oblak,
Christoph Simon, Wolfgang Sohler, Wolfgang Tittel, A solid-state memory for multiplexed quantum states of light with read-out on demand,
arXiv:1309.3202 (2013)
%
\bibitem{SokolovVasilyev} D.V. Vasilyev, I.V. Sokolov, E.S. Polzik, Phys. Rev. A 81, 020302(R) (2010)
%
\bibitem{HSQM1} T.Golubeva, Yu.Golubev, O.Mishina, A. Bramati, J. Laurat, E.Giacobino,
High speed spatially multimode atomic memory, Phys.Rev. A, 83(5), 053810 (2011)
%
\bibitem{Dantan} A.~Dantan, J.~Cviklinski, M.~Pinard, and Ph.~Grangier, Phys. Rev. A 73, 032338 (2006)
%
\bibitem{Nunn} J. Nunn, I.A. Walmsley, M.G. Raymer, K. Surmacz, F.C.
Waldermann, Z. Wang, D. Jaksch, Phys. Rev. A 75, 011401 (2007)
%
\bibitem{Lam} G. H\'{e}tet, A. Peng, M. T. Johnsson, J. J. Hope, and P. K. Lam,
Characterization of electromagnetically-induced-transparency-based continuous-variable quantum memories, Phys. Rev. A 77, 012323 (2008)
%
\bibitem{Adiabatic}  K. Samburskaya, T. Golubeva, Yu. Golubev, and E. Giacobino,
Quantum Holography upon Resonant Adiabatic Interaction of Fields with an Atomic Medium in a Lambda-Configuration, Opt. Spectr. 110(5), 775 (2011)
%
\bibitem{HSQM2} T. Golubeva, Yu. Golubev, O. Mishina, A. Bramati, J. Laurat, E. Giacobino,
High speed spatially multimode Lambda-type atomic memory with arbitrary frequency detuning, Eur. Phys. J. D, 66, 275  (2012)
%
\bibitem{Gorshkov_free} Alexey V. Gorshkov, Axel Andr\'{e}, Mikhail D. Lukin, and Anders S. Sorensen,
Photon storage in $\Lambda$-type optically dense atomic media. II. Free-space model, Phys. Rev. A 76, 033805 (2007)
%
\bibitem{Lvovsky} J. Appel, E. Figueroa, D. Korystov, M. Lobino, and A. I. Lvovsky, Phys. Rev. Lett., 100, 093602 (2008)
%
\bibitem{SPL} Golubev Yu., Golubeva T., Ivanov D.
Broadband squeezed light from phase-locked single-mode sub-poissonian lasers, Phys. Rev. A., 77, 052316 (2008)
%
\bibitem{DOPO1} Y.M. Golubev, T.Y. Golubeva, A.A. Gavrikov, C. Fabre,
Pure and mixed stats in degenerate parametric oscillation, Opt. Spectr., 106(5), 723-729 (2009)
%
\bibitem{DOPO2} V. A. Averchenko, T. Yu. Golubeva, Yu. M. Golubev, C. Fabre,
Broadband Radiation of a Degenerate Optical Parametric Oscillator Operating above Threshold in Information Applications, Opt. Spectr., 105 (5),
758-770 (2008)
%
\bibitem{Samb2012} K. S. Samburskaya, T. Yu. Golubeva, V. A. Averchenko, and Yu. M. Golubev,
Quadrature Squeezing in an Isolated Pulse of Light, Opt. Spectr., 113(1), 86-95 (2012)
%
\bibitem{Gorshkov_cavity} Alexey V. Gorshkov, Axel Andr\'{e}, Mikhail D. Lukin, and Anders S. Sorensen,
Photon storage in $\Lambda$-type optically dense atomic media. I. Cavity model, Phys. Rev. A 76, 033804 (2007)

\bibitem{Sokolov} It was proved in the talk of Ivan Sokolov on the regular seminar of Quantum Optics Lab., St.Petersburg State University (2013)

\end{thebibliography}
\end{document}